\def\year{2019}\relax
\documentclass[letterpaper]{article} 
\usepackage{aaai19}  
\usepackage{times}  
\usepackage{helvet} 
\usepackage{courier}  
\usepackage[hyphens]{url}  
\usepackage{graphicx} 
\usepackage{graphicx}  
\title{From ``Welcome New Gabbers'' to the Pittsburgh Synagogue Shooting:\\ The Evolution of Gab}
\author{Reid McIlroy-Young, Ashton Anderson\\
Department of Computer Science, University of Toronto, Canada\\
reidmcy@cs.toronto.edu, ashton@cs.toronto.edu
}

\newcommand{\xhdr}[1]{\vspace{1.7mm}\noindent{{\bf #1.}}}
\usepackage{siunitx}
\DeclareGraphicsExtensions{{.pdf, .png, .jpg}}
\graphicspath{ {./} }
\usepackage{booktabs}
\usepackage{makecell}

\begin{document}
	\maketitle
    
\begin{abstract}
		Gab, an online social media platform with very little content moderation, has recently come to prominence as an alt-right community and a haven for hate speech. We document the evolution of Gab since its inception until a Gab user carried out the most deadly attack on the Jewish community in US history. We investigate Gab language use, study how topics evolved over time, and find that the shooters' posts were among the most consistently anti-Semitic on Gab, but that hundreds of other users were even more extreme.
	\end{abstract}

\section{Introduction}

The ecosystem of online social media platforms supports a broad diversity of opinions and forms of communication. In the past few years, this has included a steep rise of alt-right rhetoric, incendiary content, and trolling behavior. In response to this, mainstream platforms have begun to take stronger stands against particularly problematic users---for example, in September, 2018, Twitter permanently banned conspiracy theorist Alex Jones from their site.

This has spurred the creation and flourishing of a social media platform called Gab, which has very few restrictions on content. Many users, either banned from other platforms or attracted to Gab's lack of moderation, have joined and contributed to Gab. It has since become a haven for members of the alt-right and has a high prevalence of hate speech and offensive content~\cite{lima2018inside,zannettou2018gab}.

On October 27, 2018, a man named Robert Bowers entered a Pittsburgh synagogue and shot 18 people. Prior to the shooting, Bowers was a Gab user and posted anti-Semitic comments, including references to the Tree of Life synagogue he attacked.

In this work, we analyze the evolution of the Gab platform since its inception until the October, 2018 shooting. How did the topics Gab users talk about change over time, and how did Gab become what it is today? We find that Gab became more entrenched in alt-right topics as time went on.

We also investigate what kind of language Gab users use. We see that alt-right topics and hate speech are considerably over-represented in Gab, whereas technical and professional topics are under-represented.

Finally, we conduct an analysis of the shooter's profile and content. After many mass shootings and terrorist attacks, analysts and commentators have often pointed out ``warning signs'', and speculate that perhaps the attacks could have been foreseen. The shooter's anti-Semitic comments and references to the synagogue he later attacked are an example of this. We compare the shooters' Gab presence with the rest of the Gab user base, and find that while he was among the most consistently anti-Semitic users, there were still hundreds of active users who were even more extreme.

\section{Related Work}

Our work draws upon three main lines of research: investigations of online polarization, work on the social media ecosystem, and studies of the Gab platform.

Since the early days of the Web, people have debated whether it would act primarily as a unifying or a dividing force. Van Alstyne and Brynjolfsson provided a useful starting point by introducing measures of cyberbalkanization and developing a model to characterize when online interaction would be balkanized or integrated~\cite{van1996electronic}. Subsequent work studied whether online spaces facilitate or hinder exposure to politically cross-cutting views~\cite{wojcieszak2009online,flaxman2016filter}.

Our work is also closely related to the large body of research on online social media platforms. Particularly relevant to our work are studies on internet culture and community-driven platforms like \textit{4chan} and \textit{Reddit}~\cite{hine2017kek,bergstrom2011don}. In this vein, researchers have discussed the relationship between platform design and the resulting potentially toxic behavior such as trolling~\cite{massanari2017gamergate,phillips2013house}.

Finally, the academic community has begun to investigate Gab, the subject of the current study. Recent work characterizes Gab's user population and finds hate speech to be prevalent~\cite{lima2018inside,zannettou2018gab}. The rapid and prominent rise of alt-right online communities has also been documented~\cite{finkelstein2018quantitative}, and the diffusion of hate speech and non-hate speech has been compared~\cite{binny2018}.

\section{Methodology}

Gab is a social media platform with similar functionality to Twitter~\cite{darroch2017}. Its nominal goal is to ``\textit{to defend free expression and individual liberty for all people}'' (Gab co-founder Andrew Torba), but is mostly used by members of the alt-right \cite{finkelstein2018quantitative}. Users can create posts, which are broadcasted to their followers, and these posts can be up-voted and down-voted by other users. Users can repost (similar to a retweet) and comment on others' posts, and the platform supports hashtags and @-mentions. Users have a unique username and can follow or be followed by other users. A defining characteristic of the service is that it has very little moderation; only illegal pornography, spam, threats, and terrorism are not allowed.



\subsection{Data Collection}

To collect the data, we started with a seed set of top users recommended by Gab to new users, and then recursively crawled users as we discovered them from the following and follower lists of users we'd already crawled. This process took six weeks to complete, and resulted in our final dataset containing 748K users and 30.5M posts (of which 15M were reposts) between August 10, 2016 and October 28, 2018.  




On October 27, 2018, Robert Bowers entered the Tree of Life synagogue in Pittsburgh and shot 11 people and injured 7 others, carrying out the deadliest attack on the Jewish community in United States history. Moments prior to the attack, Bowers, who had been active on Gab for approximately one year, posted about his intent on Gab. In the aftermath of the shooting, Gab was deplatformed and dropped by its payment provider, leading to the website going down for a week. We had collected $95\%$ of the network when this happened, and finished the crawl within a day of Gab coming back online. We consider our dataset to be representative up until October 27, 2018.


\begin{figure*}[t]
	\centering
	\includegraphics[width=\textwidth]{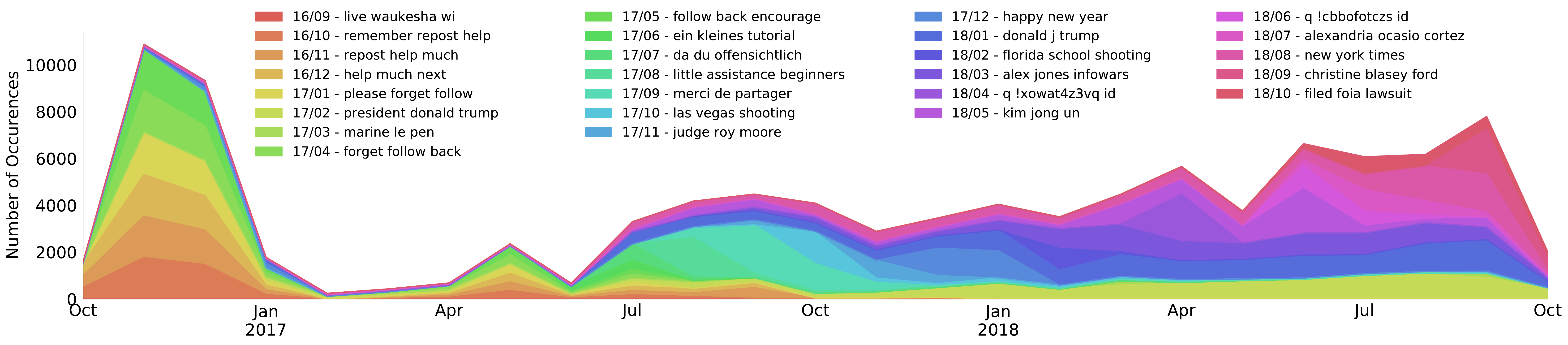}
	\caption{Topics in Gab over time. The top new trigram of each month is shown, from the site's inception to the Pittsburgh shooting. Gab started with community building, experienced an influx of non-English speakers, then became dominated by political discussion, often from an alt-right perspective.}\label{top10_flat}
\end{figure*}
\section{Results}


\subsection{Characterizing Gab Language}

We begin by characterizing the language used on Gab, and how it differs from typical online usage.

\begin{table}[t]
\tiny
	\centering
	\begin{tabular}{lrrr|lrrr}
		\toprule
		Word &  \thead{\tiny Log\\\tiny odds} &  \thead{\tiny Prob. in\\\tiny GWTWC} &  \thead{\tiny Prob. in \\\tiny Gab} & Word &  \thead{\tiny Log\\\tiny odds} &  \thead{\tiny Prob. in\\\tiny GWTWC} &  \thead{\tiny Prob. in \\\tiny Gab}   \\
\midrule
twitter         &  \num{9.8964e+00}  &  \num{5.2081e-08}  &  \num{1.0332e-03}  &  hotels          &  \num{-4.2622e+00}  &  \num{4.6846e-04}  &  \num{6.6041e-06}\\
assange         &  \num{7.4983e+00}  &  \num{2.8519e-08}  &  \num{5.1472e-05}  &  poker           &  \num{-3.6197e+00}  &  \num{2.1943e-04}  &  \num{5.8799e-06}\\
globalists      &  \num{7.4796e+00}  &  \num{4.4105e-08}  &  \num{7.8127e-05}  &  shipping        &  \num{-3.4954e+00}  &  \num{4.3617e-04}  &  \num{1.3238e-05}\\
youtube         &  \num{7.3720e+00}  &  \num{2.0597e-07}  &  \num{3.2756e-04}  &  reviews         &  \num{-3.2520e+00}  &  \num{5.2316e-04}  &  \num{2.0256e-05}\\
candidato       &  \num{7.2273e+00}  &  \num{2.9919e-08}  &  \num{4.1180e-05}  &  rating          &  \num{-3.2441e+00}  &  \num{3.5343e-04}  &  \num{1.3789e-05}\\
cannot          &  \num{7.1363e+00}  &  \num{1.5088e-07}  &  \num{1.8958e-04}  &  advertise       &  \num{-3.2438e+00}  &  \num{1.3937e-04}  &  \num{5.4385e-06}\\
facebook        &  \num{7.0799e+00}  &  \num{3.7159e-07}  &  \num{4.4119e-04}  &  software        &  \num{-3.1691e+00}  &  \num{6.3000e-04}  &  \num{2.6501e-05}\\
globalist       &  \num{6.9574e+00}  &  \num{1.3649e-07}  &  \num{1.4342e-04}  &  management      &  \num{-3.1617e+00}  &  \num{5.1724e-04}  &  \num{2.1918e-05}\\
gab             &  \num{6.8533e+00}  &  \num{1.2947e-06}  &  \num{1.2245e-03}  &  nude            &  \num{-3.1577e+00}  &  \num{1.6448e-04}  &  \num{6.9955e-06}\\
potus           &  \num{6.8443e+00}  &  \num{1.8959e-07}  &  \num{1.7790e-04}  &  mortgage        &  \num{-3.0966e+00}  &  \num{1.2610e-04}  &  \num{5.7006e-06}\\
kike            &  \num{6.7462e+00}  &  \num{6.7464e-08}  &  \num{5.7395e-05}  &  mon             &  \num{-3.0615e+00}  &  \num{1.2315e-04}  &  \num{5.7661e-06}\\
shithole        &  \num{6.6773e+00}  &  \num{7.6804e-08}  &  \num{6.0994e-05}  &  casino          &  \num{-3.0530e+00}  &  \num{1.2718e-04}  &  \num{6.0058e-06}\\
goyim           &  \num{6.6712e+00}  &  \num{6.7384e-08}  &  \num{5.3188e-05}  &  wed             &  \num{-3.0438e+00}  &  \num{1.1357e-04}  &  \num{5.4126e-06}\\
candidatos      &  \num{6.6380e+00}  &  \num{2.7039e-08}  &  \num{2.0645e-05}  &  yahoo           &  \num{-2.9969e+00}  &  \num{2.6480e-04}  &  \num{1.3227e-05}\\
tweet           &  \num{6.5067e+00}  &  \num{2.8694e-07}  &  \num{1.9211e-04}  &  newsletter      &  \num{-2.9793e+00}  &  \num{1.5274e-04}  &  \num{7.7646e-06}\\
 \bottomrule
	\end{tabular}
	\caption{The 15 most over-represented (left) and under-represented (right) words on Gab by frequency with respect to the GWTWC corpus.}\label{lodds}
\end{table}

\xhdr{Frequency analysis} In our first analysis, we compare the unigram distribution of Gab posts with the unigram distribution of a large corpus of online text derived from the Google Web Trillion Word Corpus (GWTWC), which is derived from a snapshot of webpages from 2006~\cite{brants2006web}. For each token in the Gab corpus, we measure its probability of occurring in both Gab and GWTWC and compute the log-odds of these probabilities. We examine the extremes on both sides of the log-odds spectrum to understand over- and under-represented language on Gab (see Table~\ref{lodds}). Over-represented language tends to be political (e.g. \emph{assange}, \emph{potus}) and includes anti-Semitic terms (e.g. \emph{globalists}, \emph{kike}), whereas under-represented language is mostly business terms (e.g. \emph{shipping}, \emph{management}).

To ensure that these results are robust to the corpus we're comparing to, we also computed the unigram distribution of all Reddit comments from November, 2017, and conducted the same analysis. The results are similar. The most over-represented language on Gab compared to Reddit includes the terms \emph{parkland}, \emph{moslem}, \emph{declassify}, whereas the most under-represented language is mostly related to gaming.





\xhdr{Semantic analysis} Our frequency analysis highlights the topics and terms that are disproportionately likely or unlikely to be used on Gab. But beyond frequency of occurrence, we're also interested in how the usage of particular terms on Gab differs from traditional usage. To measure this, we computed a word embedding by applying word2vec to all Gab posts, and found optimal training parameters by evaluating with analogies. This embedding captures semantic relationships between words on Gab, and places words that are used similarly close to each other in the space.  We compare this Gab embedding with the standard Google News word2vec embedding, which represents words as they are used in online news articles. This analysis allows us to understand how the semantic usage of a term on Gab compares with its usage in online news articles. 

\begin{table}[t]
\tiny
	\centering
	\begin{tabular}{lrr|lrr}
		\toprule
Word & Overlap &  Log Odds &  Word & Overlap &  Log Odds \\
\midrule
rt              &         0  &      \num{5.51}  &  smartest        &        68  &     \num{-0.04}\\
donald          &         0  &      \num{4.98}  &  harder          &        66  &     \num{-0.11}\\
q               &         0  &      \num{4.78}  &  hardest         &        65  &     \num{-0.86}\\
isis            &         0  &      \num{4.47}  &  bigger          &        62  &     \num{ 0.15}\\
dem             &         0  &      \num{4.25}  &  several         &        61  &     \num{-1.21}\\
maga            &         0  &      \num{4.13}  &  grandmother     &        61  &     \num{-0.60}\\
iq              &         0  &      \num{3.84}  &  husband         &        60  &     \num{-0.27}\\
brennan         &         0  &      \num{3.66}  &  two             &        59  &     \num{-1.25}\\
dc              &         0  &      \num{3.57}  &  strongest       &        59  &     \num{-0.91}\\
tucker          &         0  &      \num{3.52}  &  grandparents    &        58  &     \num{-0.46}\\
peter           &         0  &      \num{3.42}  &  biggest         &        58  &     \num{-0.29}\\
anon            &         0  &      \num{3.27}  &  three           &        57  &     \num{-1.96}\\
fox             &         0  &      \num{3.13}  &  disappointed    &        57  &     \num{-0.95}\\
julian          &         0  &      \num{3.10}  &  stronger        &        57  &     \num{-0.41}\\
crypto          &         0  &      \num{3.10}  &  oklahoma        &        57  &     \num{ 1.85}\\
		\bottomrule
	\end{tabular}
	\caption{Words with the lowest (left) and highest (right) semantic similarity between Gab and Google News.}\label{w2v}
\end{table}




To measure how similar the semantic meaning of a word is in Gab and Google News, we compare its $k$ nearest neighbors in the Gab embedding with its $k$ nearest neighbors in the Google News embedding, and count how large the intersection of these two sets is. To do this, we first identified all words present in both datasets, then restricted our attention to the top 10,000 of these by Gab frequency. We then ranked these words by the size of the overlap of their embedding neighborhoods, shown in Table~\ref{w2v} (with $k=100$).

\begin{table}[t]
\centering
\tiny
\begin{tabular}{lr|lr}
\toprule
    \multicolumn{2}{c}{Gab}&\multicolumn{2}{c}{Google News}\\
    \midrule
   Word &  Similarity &             Word &  Similarity \\
\midrule
        kike &            0.81 &             jews &        0.72 \\
        yid &            0.62 &           jewish &        0.64 \\
    gentile &            0.58 &     orthodox\_jew &        0.62 \\
     nigger &            0.58 &              goy &        0.62 \\
        joo &            0.57 &            hasid &        0.61 \\
        goy &            0.57 &           semite &        0.60 \\
    zionist &            0.56 &            rabbi &        0.60 \\
      rabbi &            0.56 &     orthodox\_jew &        0.59 \\
      shill &            0.55 &           jewess &        0.58 \\
       jews &            0.55 &  self\_hating\_jew &        0.58 \\
       nazi &            0.54 &         gentiles &        0.58 \\
        fag &            0.54 &           shiksa &        0.57 \\
     jewish &            0.53 &            goyim &        0.56 \\
   parasite &            0.52 &          judaism &        0.56 \\
     person &            0.51 &    sephardic\_jew &        0.56 \\
\bottomrule
\end{tabular}
\caption{Most similar terms to \textit{jew} in Gab and Google News.}\label{jewish}
\end{table}

We observe a wide variety of semantic similarities and differences. On the left side of Table~\ref{w2v} are words with completely disjoint neighborhoods in the Gab and Google News embeddings, indicating they have completely different meanings in these two contexts. Virtually all of these words are political and aligned with alt-right topics. On the right side of Table~\ref{w2v} are objective words, including family terms and numbers, with similar meanings in the two datasets.

Finally, we examine the neighborhood surrounding the term \emph{jew} in both the Gab and Google News embeddings (see Table~\ref{jewish}). The differences are striking: in Gab \emph{jew} is surrounded by offensive language and ethnic slurs, whereas on Google News it is close to mostly neutral terms, although some derogatory terms are also present.


\subsection{Topical Evolution}

Gab is associated with alt-right US politics~\cite{lima2018inside,zannettou2018gab}. How did the popular content on Gab change over time? How did the discourse evolve? A benefit of our dataset is that it is complete, as it comprises the full history of Gab since its inception. In this section, we exploit the complete nature of our dataset to address the evolution of topics on Gab over time.

To examine how Gab user's interests change over time we extracted and counted all the trigrams occurring in each original post (i.e.\ excluding reposts) after filtering out stop-words. We used trigrams because they are unambiguous, descriptive, and simple to compute. They are also a good length compromise, as they are long enough to be meaningful but not so long that they become too noisy. We filtered out trigrams that only contain @ mentions and hashtags, parts of footers that are automatically added by the platform, and spam (e.g.\ \textit{buy runescape gold}). After this cleaning, the most common trigram is ``donald j trump'', with \num{35630} occurrences. Examining the most frequent trigrams shows some of the slow patterns in topics of the community, but most of these top trigrams are relatively static, not varying in frequency much from month to month. 

Instead, a more illuminating subset of trigrams to consider are the \emph{new} popular trigrams. For each month, we select a representative trigram, which is the most frequent trigram during that month that has not previously been selected as a representative trigram. This results in a picture of the popular, current discussion topics on Gab over time, which is shown in Figure~\ref{top10_flat}. An interesting story emerges from these trigrams. In the first few months, the top topics are community-building and welcoming (e.g.\ \emph{remember repost help}, \emph{please forget follow}, and, counting reposts, \emph{welcome new gabbers}). Some are focused on contemporary events, for example \textit{live waukesha wi} in the first month is from people live posting quotations from a speech Trump gave in Waukesha, Wisconsin. Throughout the summer of 2017, people from other countries starting joining, as reflected in the French and German trigrams.  By the end of 2017 and the start of 2018, the discourse switches to alt-right political topics. While some topics, \textit{florida school shooting}, were covered by mainstream news outlets, others are more niche. For example, \textit{q !xowat4z3vq id} and \textit{q !cbbofotczs id} identify posts associated with QAnon, a far-right conspiracy theory.

\begin{figure}[t]
	\centering
	\includegraphics[width=.5\textwidth]{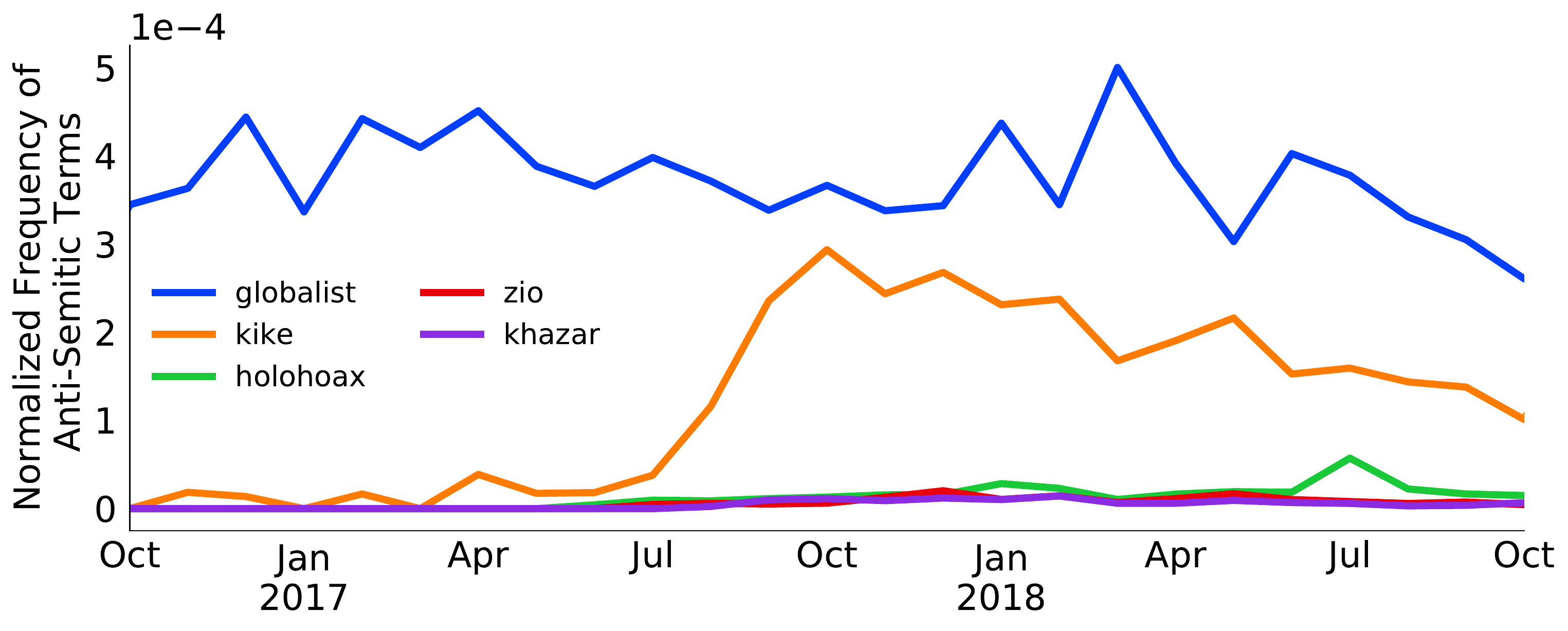}
	\caption{Normalized frequency of the top five most frequent anti-Semitic unigrams on Gab over time.}\label{antisem}
\end{figure}

As we showed in our characterization of Gab language, there is a preponderance of anti-Semitic terminology. How did anti-Semitic language evolve on Gab? To study this question, we used an external list of anti-Semitic terms and plotted their probability of appearing in posts over time (see Figure~\ref{antisem})~\cite{hatebase}. Two terms are much more common than the others: \textit{globalist}, a term with an anti-Semitic history, but which can be neutral in some contexts, and \textit{kike}, an unambiguously offensive ethnic slur. \textit{Globalist} is always the most popular term, and shows consistent usage over the course of Gab's history. However, after July, 2017, there is a striking rise of the more extreme \textit{kike}. This co-occurs with a steep rise in Gab's popularity, and the subsequent topical shift towards alt-right politics. 

\subsection{Pittsburgh Synagogue Shooting}

On October 27, 2018, Gab user Robert Bowers entered the Tree of Life synagogue and shot 18 people. Our dataset contains all of his posts except one---the one he wrote moments before carrying out the mass shooting, announcing his decision to enter the synagogue. 

After other mass shootings in the US, commentators have often pointed out ``warning signs'' in the perpetrators' pasts and speculated that the attacks were foreseeable. Here we ask: given the shooter's active presence on Gab, might we have detected something about him beforehand?

\begin{figure}[t]
	\centering
	\includegraphics[width=.5\textwidth]{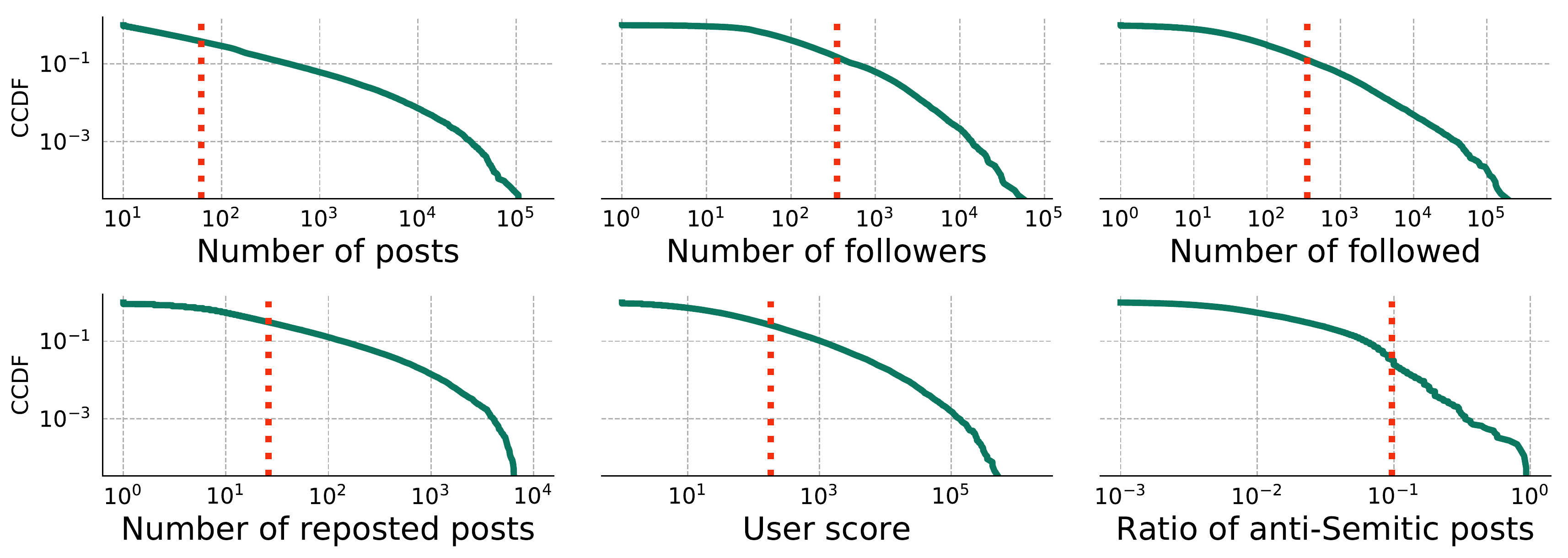}
	\caption{CCDFs of user metrics among Gab with 10 or more posts. The shooter's value is denoted by a red line.}\label{ccdfs_reduced}
\end{figure}

To examine this, we derived various metrics of his activity and compared his values to all other users with more than ten posts (73,475 active users). Figure~\ref{ccdfs_reduced} shows CCDF plots with the shooter's value denoted in red. We computed simple summary statistics, such as number of posts, the number of times they were reposted by others, the number of followers and followed, and user score to understand his activity in relation with the rest of the community. On these metrics he is unremarkable.

The main metric we computed is the ratio of a user's posts that contain anti-Semitic terms. How extreme were Bowers' posts among this extreme community? 

Bowers' posts were among the most consistently anti-Semitic on Gab. In hindsight, it may seem that the warning signs were strong. However, although Bowers was among the most extreme users, there were still hundreds of accounts with a higher ratio of anti-Semitic posts (to be precise, Bowers was 610th highest on this metric). This suggests that despite his anti-Semitic posts, prospectively identifying Bowers as a threat might not have been straightforward.




\section{Conclusion}

In this paper we examined the evolution of Gab from its inception until a violent attack was carried out by a member. We analyzed how the topics of interest to Gab users shifted from early community-building to engaging with alt-right political topics, and anti-Semitic language increased in intensity. We also considered the language used by the users and showed that in aggregate Gab users use much more offensive speech than average Internet users. We also considered if the Gab user who attacked Tree of Life synagogue was noticeably different from other users. Although he displayed ``warning signs'' by posting anti-Semitic content, there were also many users who posted even more extreme content.

\fontsize{9.0pt}{10.0pt} \selectfont
\bibliography{paper.bib}
\bibliographystyle{aaai}
\end{document}